\documentstyle[12pt]{article}
\def\l{\lambda}
\def\m{\mu}
\def\n{\nu}

\def\o{\omega}
\def\d{\delta}
\def\e{\epsilon}

\def\be{\begin{equation}}
\def\ee{\end{equation}}
\def\p{\partial}
\def\ber{\begin{eqnarray}}
\def\eer{\end{eqnarray}}
\begin{document}

\begin{center}
{\Large\bf  Gauge Theories on de Sitter space and Killing Vectors }
\vskip 1 true cm
{\bf Rabin 
Banerjee}\footnote{rabin@bose.res.in}
\vskip .8 true cm
S.N. Bose National Centre for Basic Sciences,\\
Salt Lake City, Kolkata 700098, India.\end{center}
\bigskip

\centerline{\large\bf Abstract}
\medskip

We provide a general method for studying
a manifestly covariant formulation of $p$-form  gauge
theories on the de Sitter space. This is done by stereographically projecting the
corresponding theories, 
defined on flat Minkowski space, onto the
surface of a de Sitter hyperboloid. The gauge fields in the two descriptions are mapped by conformal
Killing vectors allowing
for a very transparent analysis and compact 
presentation of results. As applications, the axial anomaly is computed and the electric-magnetic 
duality is exhibited. Finally, the zero curvature limit is shown to yield consistent results.

\newpage

\section{Introduction}
 
\bigskip

Quantum field theory on the de Sitter space time has a long history beginning from the work of Dirac \cite{dirac}.
Its popularity stemmed from the fact that it is a maximally symmetric example of a curved space time manifold.
It is a solution of the (positive) cosmological Einstein's equations having the same degree of symmetry as the flat Minkowski space time solution. Recently, however, interest in the de Sitter space has increased enormously due to physical 
consequences when it appeared to have a crucial role in the inflationary cosmological paradigm \cite{linde}. Indeed, very recently a non-zero cosmological constant has been proposed to explain the luminosity observations of the farthest supernovae
\cite{super}. The de Sitter metric will play an important role if this proposal is validated. These developments indicate
that the study of field theories on the de Sitter space deserve attention.

A standard way of constructing field theories on the de Sitter space is to use the coordinate independent
approach, also called the ambient formalism, such that there is a close resemblance with the corresponding
construction on the flat Minkowski space. For scalar fields this was initially developed in \cite{bros, bros1} which
was later extended to include gauge theories in \cite{takook, takook1}. An unpleasant feature, which
also exists in \cite{dirac}, of this approach is
that, whereas the electron wave equation involves the angular momentum operator, the gauge field equation
involves the ordinary momentum operator. Now the de Sitter space being a hyperboloid (pseudosphere), the
natural operator that should enter the equation of motion is the angular momentum operator, since  translations
on the de Sitter space correspond to `rotations' on the pseudosphere. This is generally corrected by imposing
subsidiary or homogeneity conditions on the gauge potential in order to avoid going off the hypersurface
of constant length. All these problems are absent in our approach.

Another pproach of obtaining de Sitter theories from flat space is discussed in \cite{siegel}. 
It is based on radial dimensional reduction and uses vierbein language suitable for analysing
theories on curved manifolds. Such an approach is also helpful for studying totally symmetric tensor
fields on constant curvature manifolds \cite{wally}.

In this paper we discuss a manifestly covariant  method of formulating gauge theories on the
de Sitter space which illuminates the close connection with the corresponding theories on the flat Minkowski
space. The theory on the flat space is mapped to that on the de Sitter space
by means of a stereographic projection which is basically a conformal transformation.
We show that quantities in the gauge sector (like gauge fields, field strengths etc.)
in the two descriptions are related by rules similar to usual tensor analysis, with the
conformal Killing vectors playing the role of the metric. The explicit structures of these vectors
is derived by solving the Cartan-Killing equation. Using this formalism, results for 
Yang-Mills theory on the de Sitter space are economically formulated. The equations of motion
involve the covariantised angular momentum and subsidiary conditions occurring in the usual
ambient space formalism are not needed. Also, vierbein language \cite{siegel} is not
necessary and specific properties related to de Sitter space time simplify the
technicalities considerably. The analysis is  then extended to the two form 
gauge theory. As applications of our approach we have computed the axial anomaly and also
demonstrated  electric-magnetic duality rotations in de Sitter space.

The paper is organised as follows: section 2 analyses the connection between stereographic 
projection and conformal Killing vectors, including a derivation of the latter from the
Cartan-Killing equation; section 3 treats the covariant formulation of Yang-Mills theory;
section 4 introduces the matter sector and a computation of the axial $U(1)$ anomaly is given;
section 5 reveals the electric-magnetic duality symmetry; section 6 contains an analysis of the two 
form gauge theory highlighting the appearance of a new gauge symmetry 
that does not have any analogue in the flat space; section 7 discusses the
zero curvature limit and section 8 contains a brief summary.

\section{ Stereographic Projection and Killing Vectors}

Amongst curved spacetimes, the de Sitter and anti-de Sitter spaces are the only possibilities that have
maximal symmetry admitting the highest possible number of Killing vectors. The role of these vectors
in suitably defining gauge theories on such spaces is crucial to this analysis. We shall restrict our 
discussion to de Sitter spaces only.

   The de Sitter universe is a pseudosphere in a five dimensional flat space with Cartesian coordinates
$r^a = (r^0, r^1, r^2, r^3, r^4)$ satisfying,
\be
\eta_{ab} r^a r^b = (r^0)^2 - (r^1)^2 - (r^2)^2 - (r^3)^2 - (r^4)^2 = - l^2
\label{ds1}
\ee
where $l$ is the de Sitter length parameter.
The metric of the de Sitter space $dS(4, 1)$ is induced from the pseudo Euclidean metric
$\eta = diag(+1, -1, -1, -1, -1)$. It has the pseudo orthogonal group $SO(4, 1)$ as the 
group of motions. Using the mostly negative Minkowski metric $g_{\mu\nu} = diag(+1, -1, -1, -1))$
with $\mu, \nu = 0, 1, 2, 3$, we obtain,
\be
-{1\over l^2}g_{\mu\nu} r^\mu r^\nu + (r^{'4})^2 = 1
\label{ds2}
\ee
where $r^{'4}={r^4\over l}$ is a dimensionless cordinate.

 A useful parametrisation of this space is done by exploiting the strereographic 
projection. The four dimensional stereographic coordinates $(x^\mu)$ are obtained
by projecting the de Sitter surface into a target Minkowski space. The relevant
equations are \cite{gursey},
\be
r^\mu = \Omega(x) x^\mu \,\,\,;\,\,\, \Omega(x)=\Big(1-{x^2\over 4l^2}\Big)^{-1}
\label{ds3}
\ee
and,
\be
r^{'4} = -\Omega(x) (1+{x^2\over 4l^2})
\ee
\label{ds4}
where $x^2 = g_{\mu\nu} x^\mu x^\nu$.

  The inverse transformation is given by,
\be
x^\mu = \frac{2}{1-r^{'4}} r^\mu
\label{ds5}
\ee

In order to define a gauge theory on the de Sitter space analogous stereographic
projections for gauge fields have to be obtained. This is done following the method
developed by us \cite{banerjee, banerjee1} in the example of the hypersphere. The point is that there 
is a mapping of symmetries on the flat space and the pseudosphere (e.g.translations on 
the former are rotations on the latter) that is captured by the relevant Killing vectors.
Furthermore since stereographic projection is known to be a conformal transformation, one
expects that the cherished map among gauge fields would be provided by the conformal Killing
vectors.  We may write this relation as,
 \be
 \hat A_a = K_a^\m A_\m +r_a\phi
 \label{s7}
 \ee
 where the conformal Killing vectors $K_a^\mu$ satisfy the transversality condition,
 \be
 r^a  K_a^\m = 0
 \label{s8}
 \ee
  and an additional scalar field $\phi$, which is just the normal component of $\hat A_a$, {\footnote{
Hat variables are defined on the de Sitter universe while the normal ones
are on the flat space}} is introduced,
 \be
 \phi= -\frac{1}{l^2}r^a\hat A_a
 \label{s9}
 \ee 

The five components of $\hat A$ are expressed in terms of the four components of $A$ plus a scalar degree of freedom. To simplify the analysis the scalar field is put to zero. It is straightforward to resurrect it by using the above equations. With the scalar field gone, $\hat A$ is now given by,
\be
\hat A_a = K_a^\m A_\m
\label{k}
\ee
and satisfies the transversality condition originally used by Dirac \cite{dirac}
 \be
r^a \hat A_a = 0
\label{s6}
\ee

The conformal Killing vectors $K_a^\m$ are now determined. These should satisfy the Cartan-Killing equation which, specialised to a flat four-dimensional manifold, is given by,
\be
\p^\n K_a^\m +\p^\m K_a^\n = \frac{2}{4} \p_\l K_a^\l  g^{\m\n}
\label{s10}
\ee

The most general solution for this equation  is given by \cite{FMS},
\be
K_a^\m= t_a^\m +\e_a x^\m + \omega_{a}^{\m\n} x_\n +\l_a^\m x^2 -2\lambda_a^ \sigma x_\sigma x^\m
\label{s11}
\ee
where $\o^{\m\n}= - \o^{\n\m}$.
The various transformations of the conformal group are characterised by the parameters appearing in the above equation; translations by $t$, dilitations by $\e$, rotations by $\o$ and inversions (or the special conformal transformations) by $\l$. Imposing the condition (\ref{s8}) and equating coefficients of terms with distinct powers of $x$, we find,
\ber
t_{4}^\m &=& 0 \\
x^\nu  t_\n^\m - l \o^{\m\n}_{4} x_\n - l \e_{4}x^\m &=&0 \\
-lx^2 t_{4}^\mu + 4l^2
  x^\n\e_\n x^\m + 4 l^2 x_\n\o_{\sigma}^{\m\n} x^\sigma  -4 l^3\l^\m_{4} x^2 + 8l^3 \l^\sigma_{4} x_\sigma x^\m &=& 0\\
 4l^2 x^\n\l_\n^\m x^2 - 8l^2 \l_\n^\sigma x^\n x_\sigma x^\m - l x^2\e_{4}x^\m-l \o_4^{\m\n} x_\n x^2 &=& 0\\
\l_{4}^\m x^2 - 2\l_{4}^\sigma x_\sigma x^\m &=&0
\label{s12}
\eer

Contracting the above equations (except of course the first one) by $x_\m$ leads to simplified
equations. Using them as well certain symmetry properties it is possible to obtain a solution
of the above set. The nonvanishing ones are explicitly written,
\ber
\epsilon_4 &=& l^{-1}\\
t_\n^\m &=&  g_\n^\m  \\
\l_\n^\m &=&  -\frac{1}{4l^2} g_\n^\m
\label{s23}
\eer

The basic structures of the Killing vectors, isolating the fourth component, are now written,
\be
K_\n^\m = \Big(1-\frac{x^2}{4l^2}\Big) g_\n^\m + \frac{x_\n x^\m}{2l^2} 
\label{s18a}
\ee
\be
K_{4}^\m = -K^{4\m} = \frac{x^\m}{l}
\label{s18b}
\ee

With the above solution for the Killing vectors, the stereographic projection for the gauge fields (\ref{k})
is completed leading to,
\ber
\hat A_\m &=& \Big(1-\frac{x^2}{4l^2}\Big) A_\m + \frac{x^\n x_\m}{2l^2} A_\n\\
\hat A_4 &=& \frac{x_\m}{l} A^\m
\label{newmap}
\eer

 The inverse relation is given by,
\be
\Big(1-\frac{x^2}{4l^2}\Big)A_\mu = \hat A_\mu - \frac{x_\mu \hat A_4}{2l}
\label{inverse}
\ee

Before proceeding to discuss gauge theories some properties of these Killing vectors
are summarised. 
There are two useful relations,
\be
K_a^\m K^{a\n} = \Big(1 - \frac{x^2}{4l^2}\Big)^2 g^{\m\n}
\label{k1}
\ee
and,
\be
K_a^\m K_{b\m} = \Big(1-\frac{x^2}{4l^2}\Big)^{2} \Big(g_{ab} + \frac{r_a r_b}{l^2}\Big)
\label{k2}
\ee
Note that the conformal (jacobian) factor that relates the volume element on the de Sitter space 
with that in the four-dimensional flat manifold,
\be
d^4 x = dx_0 dx_1 dx_2 dx_3 = \Big(1-\frac{x^2}{4l^2}\Big)^4 d\Omega 
\label{conformal}
\ee
naturally emerges in (\ref{k1}) and (\ref{k2}). The invariant measure is given by,
\be
d\Omega = \frac{l}{r_4} d^4r =  \Big(\frac{l}{r_4}\Big) dr_0 dr_1 dr_2 dr_3
\label{invmea}
\ee.

 Relation (\ref{k1}) shows that the product of the Killing vectors with repeated `a' indices yields, up to the conformal factor, the induced metric.  The other relation can be interpreted as the transversality condition emanating from (\ref{s8}).
For computing derivatives involving Killing vectors, a particularly useful identity is given by,
\be
K_a^\mu\p_\mu K^{a\nu} = \Big(1- \frac{x^2}{4l^2}\Big) \frac{x^\nu}{l^2}
\label{k3}
\ee

The relation (\ref{s18b}) shows that the fourth component 
is just given by the dilatation (scaling), while the other 
components (given by (\ref{s18a}) involve the special conformal transformations and the translations. 

To observe the use of these Killing vectors, 
let us analyse the generators of the infinitesimal
de Sitter transformations. In terms of the host space Cartesian coordinates $r^a$, these
are written as,
\be
L_{ab}=r_a \frac{\p}{\p r^b}- r_b \frac{\p}{\p r^a}
\label{angmom}
\ee
which satisfy the algebra,
\be
[L_{ab}, L_{cd}] = \eta_{bc} L_{ad} + \eta_{ad} L_{bc} - \eta_{bd} L_{ac} - \eta_{ac} L_{bd}
\label{angmomalg}
\ee

In terms of the stereographic coordinates the generator is expressed as,
\be
L_{ab}= (r_a K_b^\mu - r_b K_a^\mu)\p_\mu \,\,\,;\,\,\, \p_\mu = \frac{\p}{\p x^\mu}
\label{v8}
\ee
which can be put in a more illuminating form by contracting with $r^a$,
\be
r^a L_{ab} = -l^2 K_b^\mu\p_\mu
\label{new}
\ee
clearly showing how rotations on the de Sitter space are connected with the translations 
on the flat space by the Killing vectors. 

\section{Yang-Mills theory on de Sitter space}
\bigskip

In this section we discuss the formulation of Yang-Mills theory on the de Sitter space.
The theory is obtained by stereographically projecting the usual theory defined on the flat Minkowski
space. 
This is the first use of the 
generalisation effected by working in terms of the Killing vectors. 

   The pure Yang-Mills theory on the Minkowski space is governed by the standard Lagrangian,
\be
{\cal L}= -{1\over 4} Tr(F_{\mu\nu} F^{\mu\nu})
\label{v1}
\ee
where the field tensor is given by,
\be
F_{\mu\nu}=\p_\mu A_\nu - \p_\nu A_\mu -i[ A_\mu,  A_\nu ]
\label{v2}
\ee
 
 To define the field tensor on the de Sitter space we proceed systematically by looking at
the gauge symmetries.
If the ordinary potential transforms as,
\be
A_\mu' = U^{-1} (A_\mu +i\p_\mu )U
\label{gt1}
\ee
then the projected potential transforms as,
\be
\hat A_a' = K_a^\m A_\mu'=  U^{-1} (\hat A_a -\frac{i}{l^2} r^b L_{ba} )U
\label{gt2}
\ee
obtained by using (\ref{k}) and (\ref{new}).

The infinitesimal version of these transformations obtained by taking $U=e^{-i\lambda}$ is then
found to be,
\be
\delta A_\mu = D_\mu\lambda = \p_\mu \lambda -i[ A_\mu, \lambda ]
\label{v5}
\ee
 for the flat space while for the de Sitter space it is given by,
\be
\delta \hat A_a = -\frac{1}{l^2}r^b L_{ba}\lambda -i[ \hat A_a , \lambda]
\label{v10}
\ee
This is put in a more transparent form by introducing, in analogy with the flat space,
a `covariantised angular derivative' \cite{banerjee1, jackiw} on the de Sitter space,
\be
\hat{\cal L}_{ab} = L_{ab} - i [r_a \hat A_b - r_b \hat A_a,\,\,\, ]
\label{new1}
\ee
so that,
\be
\delta \hat A_a = -\frac{1}{l^2}r^b \hat{\cal L}_{ba}\lambda 
\label{new2}
\ee

The covariantised angular derivative satisfies a relation that is the covariantised version
of (\ref{new}),
\be
r^a \hat{\cal L}_{ab} = -l^2 K_b^\mu D_\mu
\label{covnew}
\ee
obtained by using the transversality condition on the gauge fields.

The field tensor $\hat F_{abc}$ on the de Sitter space is now defined. It has to be
a fully antisymmetric three index object  that transforms covariantly. The
covariantised angular derivative is the natural choice for constructing it. We define,
\be
\hat F_{abc} = \Big( L_{ab}\hat A_c -ir_a [\hat A_b, \hat A_c]\Big) + c.p.
\label{v12}
\ee
where $c.p.$ stands for the other pair of terms involving cyclic permutations in $a, b, c$.

To see that $F_{abc}$ transforms covariantly it is convenient to recast this in a form involving the Killing vectors,
analogous to the relation (\ref{k}). Indeed it is mapped to the field tensor on the flat space by the following
relation,
\be
\hat F_{abc}= \Big(r_a K_b^\mu K_c^\nu +r_b K_c^\mu K_a^\nu + r_c K_a^\mu K_b^\nu\Big)F_{\mu\nu}
\label{v11}
\ee
so that symmetry properties under exchange of the indices is correctly preserved.
To show the equivalence, (\ref{k}) and (\ref{v8}) are used to simplify (\ref{v12}),
yielding,
\be
\hat F_{abc} = \Big(r_a K_b^\mu - r_b K_a^\mu\Big)\p_\mu\Big(K_c^\nu A_\nu\Big)
-i r_a \Big[K_b^\nu A_\nu, K_c^\mu A_\mu \Big] + c.p.
\label{v13}
\ee
The derivatives acting on the Killing vectors sum up to zero on account of the identity,
\be
\Big(r_a K_b^\mu - r_b K_a^\mu\Big)\p_\mu K_c^\nu  +c.p. = 0
\label{v14}
\ee
The derivatives acting on the potentials, together with the other pieces, combine to
reproduce (\ref{v11}), thereby completing the proof of the equivalence.

  It is now trivial to see, using the above relation (\ref{v11}), that $F_{abc}$ transforms
covariantly simply because $F_{\mu\nu}$ does. The action for the Yang-Mills theory on the
de Sitter space is defined by first considering the repeated product of the field tensors.
Taking (\ref{v11}) and using the
transversality of the Killing vectors, we get,
\be
\hat F_{abc}\hat F^{abc}=- 3l^2\Big(K^{b\mu} K^{c\nu} K_b^\lambda K_c^\rho\Big)F_{\mu\nu} F_{\lambda\rho}
\label{v18}
\ee
Finally, using (\ref{k1}), we obtain,
\be
\hat F_{abc}\hat F^{abc}= -3l^2 \Big(1-\frac{x^2}{4l^2}\Big)^4 F_{\mu\nu} F^{\mu\nu}
\label{v19}
\ee
Using this identification as well as (\ref{conformal}),
the actions on the flat space and the de Sitter space are mapped as,
\be
S =  -{1\over 4}\int d^4x F_{\mu\nu} F^{\mu\nu} = {1\over {12 l^2}}\int d\Omega\hat F_{abc}\hat F^{abc}
\label{v20}
\ee
The lagrangian following from this action is given by,
\be
{\cal L}_\Omega ={1\over {12 l^2}}\hat F_{abc}\hat F^{abc}
 \label{v21}
\ee

This completes the construction of  the Yang-Mills action which can be taken as the starting point for
 calculations  on the de Sitter
space.

\section {The U(1) Axial  Anomaly}

\bigskip

The inclusion of the matter sector is also conveniently done with the help of the Killing vectors. 
Form invariance of the interaction requires that,
\be
\int dx (j_\mu A^\mu) = \int d\Omega (\hat j_a \hat A^a)
\label{m1}
\ee
where $j_\mu$ and $\hat j_a$ are the currents in the two decsriptions.
It is clear therefore that the currents are also mapped by a relation similar to
(\ref{k}). However since the measure is given by (\ref{conformal}), the currents will
involve the conformal factor. 

A simple calculation shows that, to satisfy (\ref{m1}), the desired map is provided by,
\be
 \hat j_a =\Big(1-\frac{x^2}{4l^2}\Big)^2 K_a^\m j_\m 
 \label{m3}
 \ee
Naturally the current on the de Sitter space also satisfies the transversality condition
which is taken in the literature \cite{dirac, bros, takook},
\be
r^a \hat j_a = 0
\label{current}
\ee

As an application of this formulation it is possible to compute the anomaly on the de Sitter
space from a knowledge of the corresponding expression on the flat space. To simplify matters
and to avoid a cluttering of notations, we concentrate on the abelian Adler-Bell-Jackiw anomaly.
For the axial current, employing a gauge invariant regularisation, the familiar result on the flat space
is known to be,
\be
\p_\mu j^{\mu 5} = {1\over {16\pi^2}}\e_{\mu\nu\lambda\rho}F^{\mu\nu}F^{\lambda\rho}
\label{anomaly4}
\ee

Using (\ref{new}) and the definition of the current (\ref{m3}) (appropriately 
interpreted for the axial vector currents), it is possible to obtain the
identification,
\be
 r^a L_{ab} \hat j^{b5} = - l^2 \Big(1-\frac{x^2}{4l^2}\Big)^4 \p_\mu j^{\mu 5}
\label{m9}
\ee 
In getting at the final result, use was made of the identity
(\ref{k3}). 

This provides a map for one side of (\ref{anomaly4}). To obtain an analogous form for the other side,
 it is necessary to consider the completely antisymmetric
tensor $\e_{\mu\nu\lambda\rho}$ whose value is the same in all systems. 

In order to provide a mapping among the $\epsilon$-tensors in the two spaces, we adopt the same rule (\ref{v11}) used for defining the antisymmetric field tensor.
However there is a slight subtlety. Strictly speaking, this Levi-Civita epsilon is a tensor density.
Hence its transformation law is modified by an appropriate conformal (weight) factor,
\be
\e_{abcde}=\frac{1}{l} \Big(1 - \frac{x^2}{4l^2}\Big)^{-4}  
\Big(r_a K_b^\mu K_c^\nu K_d^\lambda K_e^\rho + cyclic\,\,\,\, permutations\,\,\,in\,\,\,(a, b, c, d, e) \Big)\e_{\mu\nu\lambda\rho}
\label{m5}
\ee
It is possible to verify the above relation by an explicit calculation, taking the convention
that both the epsilons are $+1 (-1)$ for any even (odd) permutation of distinct entries $(0,1, 2, 3, 4 )$ in that order.

Now the explicit expressions for the anomaly are  identified with the minimum of effort.
Indeed, using (\ref{v11}) and (\ref{m5}), it follows that,
\be
r_a\e_{bcdef}\hat F^{abc}\hat F^{def}= 
3 l^3 \Big(1 - \frac{x^2}{4 l^2}\Big)^4\e_{\mu\nu\lambda\rho}F^{\mu\nu}F^{\lambda\rho}
\label{m11}
\ee

The weight factors cancel out from both sides of the 
anomaly equation and we obtain,
\be
 r_a L^{ab} \hat j_{b5} = - {1\over {48\pi^2 l}}r_a\e_{bcdef}\hat F^{abc}\hat F^{def}
\label{m12}
\ee

This is the desired anomalous current divergence equation in the de Sitter space. It is 
the exact analogue of the ABJ-anomaly equation in the flat space.

There is another way in which the anomaly equation can be expressed. To see this consider
the product of the operator on one side of (\ref{m12}) with the radius vector to get,
\be
r_c r_a L^{ab} \hat j_{b5} = -l^2\Big(L_{cb} + r_b K_c^\mu \partial_\mu\Big)\hat j^{b5}
\label{m12a}
\ee
where (\ref{new}) has been used. The second term on the right hand side is simplified by
exploiting (\ref{m3}) and the identity,
\be
r^b K_c^\mu \partial_\mu K_b^\sigma = -K_c^\sigma
\label{m12b}
\ee
to yield,
\be
r_c r_a L^{ab} \hat j_{b5} = l^2\Big( \hat j_{c5} - L_{cb}\hat j^{b5}\Big)
\label{m12c}
\ee
Thus the anomaly equation (\ref{m12}) takes the form,
\be
\hat j_{g5} -  L_{gb} \hat j^{b5} = - {1\over {48\pi^2 l^3}}r_g r_a\e_{bcdef}\hat F^{abc}\hat F^{def}
\label{m12d}
\ee
Compatibility between the two forms (\ref{m12}) and (\ref{m12d}) is easily established by
contracting the latter with $r^g$ and using the transversality (\ref{current}) of the current.

  The normal Ward identity for the vector current is obtained by setting the right hand side of 
either (\ref{m12}) or (\ref{m12d}) equal to zero.

\section{Duality Symmetry}

The well known electric-magnetic duality symmetry swapping 
field equations with the Bianchi identity in flat space has an exact counterpart on
the de Sitter hyperboloid. To see this it is essential to introduce the dual field 
tensor that enters the Bianchi identity. The dual tensor is defined by,
\be
\tilde F_{ab} = -\frac{1}{6}\epsilon_{abcde}\hat F^{cde}
\label{d1}
\ee

Using (\ref{v11}) and (\ref{m5}) together with the properties of the Killing vectors the 
dual on the de Sitter space is expressed in terms of the dual on the flat space as,
\be
\tilde F_{ab} = l K_a^\lambda K_b^\rho \tilde F_{\lambda\rho}
\label{d2}
\ee
where the flat space dual is given by,
\be
\tilde F_{\lambda\rho} = \frac{1}{2}\epsilon_{\lambda\rho\mu\nu} F^{\mu\nu}
\label{d3}
\ee

The Bianchi identity is then given by,
\be
r_a L^{ab} \tilde F_{bc} = 0
\label{d4}
\ee
This is confirmed by a direct calculation. Alternatively, it becomes transparent by projecting it
on the flat space by means of Killing vectors. Using the basic definitions and the
identity,
\be
K_b^\rho \p_\rho(K^{b\mu} K_c^\nu)\tilde F_{\mu\nu} = 0
\label{zc2}
\ee
we obtain,
\be
r_a L^{ab} \tilde F_{bc} = -l^3 K^{b\mu}K_b^\lambda K_c^\rho \p_\mu\tilde F_{\lambda\rho}
\label{d5}
\ee
Finally, exploiting  (\ref{k1}) we get the desired projection,
\be
r_a L^{ab} \tilde F_{bc} = -l^3 \Big(1-\frac{x^2}{4l^2}\Big)^2 K_c^\rho \p^\lambda\tilde F_{\lambda\rho}
\label{d6}
\ee
which vanishes since $\p^\lambda\tilde F_{\lambda\rho}=0$.

Now the abelian equation of motion following from a variation of the action (\ref{v20})
is given by,
\be
L_{ab} \hat F^{abc} = 0
\label{d7}
\ee

The duality transformation is next discussed. Analogous to the flat space rule, 
$\tilde F\rightarrow F; F\rightarrow -\tilde F$ the duality map here is provided by,
\be
\tilde F_{ab} \rightarrow \frac{r^c}{l} \hat F_{abc}\ \ \ ;\ \ \ \frac{r^c}{l}\hat F_{abc}\rightarrow -\tilde F_{ab}
\label{d8}
\ee
It is easy to check the consistency of this map. The inverse of (\ref{d1}) yields,
\be
\hat F_{abc} = -\frac{1}{2}\epsilon_{abcde}\tilde F^{de}
\label{d9}
\ee
Under the first of the maps in (\ref{d8}), the above relation is transformed as,
\be
\hat F_{abc}\rightarrow \frac{1}{l}\Big(r_a \tilde F_{bc} +r_b \tilde F_{ca}+ r_c \tilde F_{ab}\Big)
\label{d10}
\ee
where use was made of (\ref{d9}) at an intermediate step. Contracting the above map by $r^c$
immediately leads to the second relation in (\ref{d8}).

Now the effect of the duality map on the equation of motion (\ref{d7}) is considered. Using 
(\ref{d10}) and the correspondence (\ref{d2}) along with the identity (\ref{zc2}) we find,
\be
\frac{1}{2}L_{ab} \hat F^{abc} \rightarrow -l^2\Big(1-\frac{x^2}{4l^2}\Big)^2 K^{c\rho}\p^\mu \tilde F_{\mu\rho}
\label{d11}
\ee
Finally, using (\ref{d6}) we obtain the cherished mapping,
\be
\frac{1}{2}L_{ab} \hat F^{abc} \rightarrow \frac{1}{l} r_a L^{ab} \tilde F_{bc}
\label{d11}
\ee
showing how the equation of motion passes over to the Bianchi identity. Likewise the 
other map swaps the Bianchi identity to the equation of motion.

It is feasible to perform a continuous $SO(2)$ duality rotations through an angle $\theta$. The relevant transformations are
then given by,
\ber
\frac{r^c}{l} F_{abc}' &=& \mbox{Cos} \theta \ \ \frac{r^c}{l} F_{abc} - \mbox{Sin} \theta\ \ \tilde F_{ab} \\
\tilde F_{ab}' &=& \mbox{Sin} \theta \ \ \frac{r^c}{l} F_{abc} + \mbox{Cos} \theta \ \ \tilde F_{ab}
\label{d12}
\eer
The discrete duality transformation corresponds to $\theta=\frac{\pi}{2}$.

\section{ Formulation of Antisymmetric Tensor Gauge Theory}

\bigskip

The general formalism developed so far is particularly suited for obtaining a 
 formulation of $p$-form gauge theories. Here we discuss it for the 
second rank antisymmetric tensor gauge theory. Also, there are some features
which distinguish it from the analysis for the  vector gauge theory.
The extension for higher forms 
is obvious. Both abelian and nonabelian theories will be considered. To set
up the formulation it is convenient to begin with the abelian case which can
be subsequently generalised to the nonabelian version. The action for a free
2-form gauge theory in flat four-dimensional  Minkowski space is given by \cite{KR},
\be
S= -{1\over {12}}\int d^4 x F^{\mu\nu\rho}F_{\mu\nu\rho}
\label{t1}
\ee
where the field strength is defined in terms of the basic field as,
\be
 F_{\mu\nu\rho}= \p_\mu B_{\nu\rho}+ \p_\nu B_{\rho\mu}+ \p_\rho B_{\mu\nu}
\label{t2}
\ee
The infinitesimal gauge symmetry is given by the transformation,
\be
\d B_{\mu\nu} = \p_\mu \Lambda_\nu -\p_\nu \Lambda_\mu
\label{t3}
\ee
which is reducible since it trivialises for the choice $\Lambda_\mu = \p_\mu\lambda$.

It is sometimes useful to express the action (or the lagrangian) in a first order form
by introducing an extra field,
\be
{\cal L}= -{1\over {8}}\e_{\mu\nu\rho\sigma} F^{\mu\nu}B^{\rho\sigma}+ {1\over 8}A^\mu A_\mu
\label{t4}
\ee
where the $B\wedge F$ term involves the field tensor,
\be
F_{\mu\nu}=\p_\mu A_\nu - \p_\nu A_\mu
\label{t5}
\ee
Eliminating the auxiliary $A_\mu$ field by using its equation of motion, the previous
form (\ref{t1}) is reproduced. The gauge symmetry is given by (\ref{t3}) together with
$\d A_\mu=0$. The first order form is ideal for analysing the nonabelian theory.

To express the theory on the de Sitter pseudosphere, the mapping of the tensor field is first given.
From the previous analysis, it is simply given by,
\be
\hat B_{ab} = K_a^\mu K_b^\nu B_{\mu\nu} 
\label{t9}
\ee
The tensor field with the latin indices is defined on the pseudosphere while those with
the greek symbols are the usual one on the flat space.
This is written in component notation by using the explicit form  for the Killing vectors 
given in (\ref{s18a}) and (\ref{s18b}),
\be
\hat B_{\m\n} =\Big(1 -\frac{x^2}{4l^2}\Big)\Big((1-\frac{x^2}{4l^2}) B_{\m\n} + \frac{x^\rho x_\n}{2l^2} B_{\m\rho}
+\frac{x^\rho x_\m}{2l^2} B_{\rho\n}\Big)
\label{t10}
\ee
and,
\be
\hat B_{\mu 4} = \frac{1}{l}\Big(1-\frac{x^2}{4l^2}\Big) x^\rho B_{\m\rho}
\label{t11}
\ee
These are the analogues of (\ref{newmap}). The inverse relation is given by,
\be
\Big(1-\frac{x^2}{4l^2}\Big)^4 B^{\m\n} = K_a^\m K_b^\n \hat B^{ab}
\label{t12}
\ee
which may also be put in the form,
\be
\Big(1-\frac{x^2}{4l^2}\Big)^2 B_{\m\n} =  \hat B_{\m\n} +\frac{x_\m \hat B_{\n 4}}{2l}- \frac{x_\n \hat B_{\m 4}}{2l}
\label{t13}
\ee
which is the direct analogue of (\ref{inverse}).

Next, the gauge transformations are discussed. From (\ref{t3}), the defining relation (\ref{t9}) and the 
angular momentum operator (\ref{v8}), infinitesimal transformations are given by,
\be
\d \hat B_{ab}= -\frac{r^c}{l^2}\Big(K_b^\m L_{ca} -K_a^\m L_{cb}\Big)\Lambda_\m
\label{t14}
\ee
In this form the expression is not manifestly covariant. This may be contrasted with
(\ref{v10}) which has this desirable feature. The point is that an appropriate map
of the gauge parameter is necessary. In the previous example the gauge parameter was a scalar
which retained its form. Here, since it is a vector, the required map is provided by a relation
like (\ref{k}), so that,
\be
\hat\Lambda_a = K_a^\mu \Lambda_\mu
\label{t14.1}
\ee
Pushing the Killing vectors through the angular momentum operator and using the above 
map yields, after some simplifications,
\be
\d \hat B_{ab}= -\frac{1}{l^2}\Big[ r^c\Big( L_{ca}\hat\Lambda_b - L_{cb}\hat\Lambda_a \Big)
- r_a\hat\Lambda_b + r_b\hat\Lambda_a\Big]
\label{t14.3}
\ee

It is also reassuring to note that (\ref{t14.3}) manifests the reducibility of the
gauge transformations. Since $\Lambda_\mu=\p_\mu\lambda$ leads to a trivial gauge
transformation in 
flat space, it follows from (\ref{t14.1}) that the corresponding feature should be present
in the pseudospherical formulation when,
\be
\hat\Lambda_a =  r^c L_{ca}\lambda
\label{t14.5}
\ee
It is easy to check that with this choice, the gauge transformation (\ref{t14.3}) 
trivialises; i.e. $\d \hat B_{ab}=0$.

The field tensor on the pseudosphere is constructed from the usual one given in (\ref{t2}).
Since the Killing vectors play the role of the metric in connecting the two surfaces, this
expression is given by a natural extension of (\ref{v11}),
\be
\hat F_{abcd}= \Big(r_a K_b^\mu K_c^\nu K_d^\rho+r_b K_c^\mu K_a^\nu K_d^\rho + r_c K_d^\mu K_a^\nu
K_b^\rho +r_d K_a^\mu K_c^\nu K_b^\rho\Big)F_{\mu\nu\rho}
\label{t15}
\ee
Note that cyclic permutations have to taken carefully since there is an even number of
indices.

In terms of the basic variables, the field tensor is expressed as,
\be
\hat F_{abcd}= \Big(L_{ab}\hat B_{cd} +L_{bc}\hat B_{ad} + L_{bd}\hat B_{ca} + L_{ca}\hat B_{bd} +
L_{da}\hat B_{cb}+
L_{cd}\hat B_{ab}\Big)
\label{t16}
\ee

To show that (\ref{t15}) is 
equivalent to (\ref{t16}), the same strategy as before, is 
adopted. Using the definition of the angular 
momentum (\ref{v8}), (\ref{t16}) is simplified
as,
\be
\hat F_{abcd}=\Big(r_a K_b^\mu - r_b K_a^\mu\Big)\p_\mu\Big( K_c^\nu K_d^\sigma B_{\nu\sigma} \Big)
+............
\label{t17}
\ee
where the carets denote the inclusion of other similar 
(cyclically permuted) terms. Now there are two types
of contributions. Those where the derivatives act on the Killing vectors and those where
they act on the fields. The first class of terms cancel out as a consequence of an identity
that is an extension of (\ref{v14}). The other class combines to reproduce (\ref{t15}).

The action on the de Sitter pseudosphere is now obtained by first taking a repeated product of the field
tensor (\ref{t15}). Using the properties of the Killing vectors, this yields,
\be
\hat F_{abcd}\hat F^{abcd}= -4l^2\Big(1-\frac{x^2}{4l^2}\Big)^6 F_{\mu\nu\rho}F^{\mu\nu\rho}
\label{t18}
\ee
From the definition of the flat space action (\ref{t1}) and the volume element (\ref{conformal}),
it follows that the above identification leads to the pseudospherical action,
\be
S_\Omega =\frac{1}{48 l^2}\int d\Omega \Big(1-\frac{x^2}{4l^2}\Big)^{-2}\hat F_{abcd}\hat F^{abcd}
\label{t19}
\ee

Thus, up to a conformal factor, the corresponding lagrangian is given by,
\be
{\cal L}_\Omega =\frac{1}{48 l^2}\hat F_{abcd}\hat F^{abcd}
\label{t20}
\ee

By its very construction this lagrangian would be invariant under the gauge transformation
(\ref{t14.3}). There is however another type of gauge symmetry 
which does not seem to have any analogue in the flat space. 
To envisage such a  possibility, consider a  transformation of the
type {\footnote{Recently such a transformation was considered on the 
hypersphere \cite{M, banerjee}}},
\be
\delta \hat B_{ab}=L_{ab}\lambda
\label{t21}
\ee
which  could be a meaningful gauge symmetry
operation on the de Sitter space. However, in flat space, it
leads to a trivial gauge transformation. To see this explicitly, consider the effect of
(\ref{t21}) on (\ref{t12}),
\be
\Big(1-\frac{x^2}{4l^2}\Big)^4 \delta B^{\m\n} = K_a^\m K_b^\n L_{ab}\lambda
\label{t22}
\ee
Inserting the expression for the angular momentum from (\ref{v8}) and
exploiting the transversality (\ref{s8}) of the Killing vectors, it follows that,
\be
 \delta B_{\m\n} = 0
\label{t23}
\ee
thereby proving  the statement. To reveal that (\ref{t21})
indeed leaves the lagrangian (\ref{t20}) invariant, it is desirable 
to recast it in the form,
\be
{\cal L}_\Omega =\frac{1}{32}\hat \Sigma_{a}\hat \Sigma^{a}
\label{t24}
\ee
where,
\be
\hat \Sigma_a = \e_{abcde}L^{bc}\hat B^{de}
\label{t25}
\ee
Under the gauge transformation (\ref{t21}), a simple algebra shows that $\delta \hat \Sigma_a = 0 $ and hence 
the lagrangian remains invariant.

The inclusion of a nonabelian gauge group is feasible. Results follow logically
from the abelian theory with suitable insertion of the nonabelian indices. As
remarked earlier it is useful to consider the first order form (\ref{t4}). The
lagrangian is given by its straightforward generalisation \cite{FT},
where the nonabelian field strength has already been defined in (\ref{v2}). It is
gauge invariant under the nonabelian generalisation of (\ref{t3}) with the 
ordinary derivatives replaced by the covariant derivatives with respect to the potential 
$A_\mu$, and $\d A_\mu = 0$. By the help of our equations it is possible to
project this lagrangian on the de Sitter space. For instance, the corresponding gauge 
transformations look like,
\be
\d \hat B_{ab} = -\frac{1}{l^2} \Big[r^c\Big( L_{ca}\hat\Lambda_b - L_{cb}\hat\Lambda_a \Big)
- r_a\hat\Lambda_b + r_b\hat\Lambda_a\Big] + [\hat A_a ,  \hat\Lambda_b]
\label{t27}
\ee
and so on.

Matter fields may be likewise defined. The fermion current $j_{\mu\nu}$ will
be defined just as the two form field, except that conformal weight factors appear,
so that form invariance of the interaction is preserved,
\be
\int dx (j_{\mu\nu} B^{\mu\nu}) = \int d\Omega (\hat j_{ab} \hat B^{ab})
\label{t28}
\ee
quite akin to (\ref{m1}).

\section {Zero Curvature Limit}
\bigskip

The null curvature limit (which is also equivalent to a vanishing cosmological constant) is obtained
by setting $l\rightarrow \infty$. Then the de Sitter group contracts to the 
Poincare group so that the field theory on the de Sitter space should contract to the corresponding theory
on the flat Minkwski space. This  is shown  very conveniently in the present formalism 
using Killing vectors. The example of Yang Mills theory with sources will be considered.

The equation of motion in the de Sitter space  obtained by varying the action composed of the pieces
(\ref{v20}) and (\ref{m1}) is found to be,
\be
\frac{1}{2 l^2}\hat{\cal L}_{ab}\hat F^{abc} - \hat j^c = 0
\label{zc1}
\ee

The operator appearing in the above equation is now mapped to the flat space. The mapping
for the usual angular
momentum part is first derived,
\be
L_{ab} \hat F^{abc} = 2 r_a K_b^\mu \p_\mu\Big([r^a K^{b\nu} K^{c \rho} +c.p.] F_{\nu\rho}\Big)
\label{zero}
\ee
Using the transversality condition and the identities among the Killing vectors it is seen that the
only nonvanishing contribution comes from the action of the derivative on the field tensor yielding,
\be
 L_{ab} \hat F^{abc} = -2 l^2\Big(1-\frac{x^2}{4l^2}\Big)^2 K^{c\rho} \p^\mu F_{\mu\rho}
\label{zero1}
\ee
 It is straightforward to generalise this for
the covariantised angular momentum and one finds,
\be
\hat {\cal L}_{ab} \hat F^{abc} = -2 l^2\Big(1-\frac{x^2}{4l^2}\Big)^2 K^{c\rho} D^\mu F_{\mu\rho}
\label{zc4}
\ee

Using the map (\ref{m3}) for the currents, the equation of motion on the de Sitter space finally gets
projected on the flat space as,
\be
\Big(1-\frac{x^2}{4l^2}\Big)^2 K^{c\rho}\Big( D^\mu F_{\mu\rho} +\hat j_\rho\Big) = 0
\label{zc5}
\ee
This equation is now multiplied by the Killing vector $K_c^\lambda$. Using the identity among the Killing
vectors yields,
\be
\Big(1-\frac{x^2}{4l^2}\Big)^4 \Big( D^\mu F_{\mu\lambda} +\hat j_\lambda\Big) = 0
\label{zc6}
\ee
 
The zero curvature limit $(l\rightarrow \infty)$ is now taken. The prefactor simplifies to unity and the standard flat
space Yang Mills equation
with sources is reproduced.

\section{Discussions}

\bigskip

We have provided a manifestly  covariant formulation of vector and tensor
gauge theories on the de Sitter hyperboloid. It was done  by mapping  the usual forms of these theories, defined
on the Minkowski flat surface, onto the hyperboloid by the method of stereographic
projection. A distinctive feature was the abstraction of the relevant conformal
Killing vectors by solving the Cartan-Killing equation. The importance of these
Killing vectors lay in the fact that tensor forms constructed by taking their 
products acted like a metric connecting the results between the flat space
and the hyperboloid. This essentially new ingredient was crucial for generalisations
to include higher form gauge theories. We feel that the present prescription relates 
theories in flat space time with those in de Sitter space time compactly and elegantly. 
All technicalities are reduced to simple algebraic properties of the Killing vectors.
Although our analysis was shown for
the two form case, extension to any $p$-form gauge theory is clear. Also, the
present analysis is easily applicable for completely symmetric tensor fields
as well as for arbitrary dimensions with obvious changes. These changes entail trivial modifications in 
the properties of the Killing vectors.

An advantage of this approach is that derivatives like $\frac{\p}{\p r^a}$ always occur as
the angular momentum $L_{ab}=r_a\p_b - r_b\p_a$ in the electromagnetic field strengths
or in the equations of motion. In the ambient space approach \cite{dirac, bros, bros1, takook} 
both types of (linear and angular) derivatives occur. Consequently homogeneity
conditions \cite{dirac, bros, bros1, takook} are  required in order to avoid going off the 
hypersurface of constant $r^2$. This is completely avoided in our analysis.

  The expressions throughout this paper have the most simple and natural Minkowskian-type structures
for obvious reasons. Indeed a chunk of the results on the hypersphere done by us \cite{banerjee,
banerjee1} could be appropriately manipulated to yield reults on the de Sitter space. This was
feasible since the de Sitter space is essentially a pseudosphere that is related to the hypersphere
by a Wick-like rotation. The possibility that a hyperspherical analysis could be useful for a de Sitter
analysis was already mooted by Adler \cite{adler}, though it was in a very restrictive sense valid for
abelian theories and without the powerful use of Killing vectors. 

As applications we have provided new results for the axial anomaly as well as shown the existence
of  electric-magnetic duality rotations on the de Sitter hyperboloid. Both these phenomenon were obtained
by a direct mapping of the known results   on the flat Minkowski space. Indeed we feel that our 
approach gives an intuitive understanding of the closeness of the formulation of gauge field theories
on  flat and de Sitter spaces. Incidentally, formulating a gauge theory on the de Sitter space that
mimics the corresponding formulation on the flat space has been an important objective of several
authors \cite{dirac, bros, bros1, takook, siegel, wally}.

Finally the zero curvature $(l\rightarrow \infty)$
limit was discussed. The results of de Sitter space  gauge theory expectedly
contracted to the Minkowski space gauge theory thereby vindicating our analysis. In cosmological terms this limit is
essentially the zero cosmological constant limit. The other extreme $(l\rightarrow 0)$ corresponds to the 
infinite cosmological constant limit. In this case the de Sitter space tends to the conic spacetime \cite{ald}.
It would be interesting to construct a gauge theory on this conic space and see whether the results given here
pass on to that theory in the $(l\rightarrow 0)$ limit.

\end{document}